\documentstyle[amssymb,aps]{revtex}
\input{psfig.tex}

\newcommand{\nc}{\newcommand}
\nc{\be}{\begin{eqnarray}}
\nc{\ee}{\end{eqnarray}}
\nc{\ch}{$\chi^2$}
\nc{\pk}{PLANCK}
\nc{\cl}{$C_l$}
\nc{\ct}{$C_{Tl}$}
\nc{\lam}{$\Lambda$CDM}
\nc{\ma}{Model 1}
\nc{\mb}{Model 2}

\begin{document}

\draft
\title{Can inflationary models of cosmic perturbations evade the
secondary oscillation test?} 
\author{Alex Lewin$^{1,2}$ and Andreas Albrecht$^{1}$}
\address{$^{1}$Department of Physics\\
One Shields Ave.\\
University of California\\
Davis, CA 95616\\
$^{2}$The Blackett Laboratory, Imperial College, \\
Prince Consort Road, London SW7 2BZ, UK}

\maketitle
\begin{abstract}
We consider the consequences of an observed Cosmic Microwave
Background (CMB) temperature anisotropy spectrum containing no
secondary oscillations. While  such a spectrum is generally considered
to be a robust signature of active structure formation,  
we show that such a spectrum {\em can} be produced by (very unusual) 
inflationary models or other passive evolution models.  However, we
show that for  
all these passive models the characteristic oscillations would show up in 
other observable spectra. Our work shows that when 
CMB polarization and matter power spectra are taken into account
secondary oscillations are indeed a signature of even these very
exotic passive models. 
We construct a measure of the observability 
of secondary oscillations in a given experiment, and show that even
with foregrounds both the MAP and \pk\ satellites should be able to 
distinguish between models with and without oscillations.  Thus we
conclude that inflationary and other passive models can {\em not}
evade the secondary oscillation test.
\end{abstract}


\section{Introduction}\label{sec.intro}

Secondary oscillations in the angular power spectrum of CMB anisotropies are 
broadly understood to be a signature of inflationary models for the
origin of the primordial perturbations.  The inflationary models are part of a 
wider class of `passive' models, where the perturbations are set up
at very early times, and then evolve according to linear perturbation
theory for many  orders of magnitude of expansion.  

In linear perturbation 
theory, the physics of gravitational collapse (on scales greater than the 
Jeans length) results in a `squeezing' effect as the presence of one 
growing and one decaying solution draws any initial conditions into 
a prefered region of phase space.  On scales below the Jeans 
length there is oscillatory behavior at early times, but in passive
models each mode has already experience a long period of squeezing by
the time the oscillatory behavior kicks in.  The region of phase space
prefered by the squeezing corresponds to a fixed phase of oscillation in the
oscillatory epoch.  The fact that different wavenumbers have their
oscillations fixed at different phases ultimately results in the
characteristic `secondary oscillations' in the CMB angular power
spectrum\cite{HS1,HS2,hw}.

The active models are fundamentally different\cite{acfm}.  The
canonical examples are the cosmic defect models. 
The key difference is that these models 
have a matter component (e.g. cosmic defects) which is
evolving in a non-linear manner from very early times.  One can think
of the non-linear 
component as a source which constantly seeds pertubations in the other
matter.  This kind of non-linear `force term' disrupts the
squeezing effects which are present at the linear level, and those
realistic defect models for which calculations have been performed
have shown no secondary oscillations whatsoever in the CMB anisotropy
power\cite{abr,ACDKSS,PenEtal:97}. 
									
Thus the detection, or otherwise, of these acoustic oscillations is 
cited as a strong indication of whether or not the evolution of 
the perturbations was active or passive.  In fact, the absence of
these oscillations is regarded as one of the few ways that aspects of
inflationary cosmology can be falsified\cite{coher}. But what if 
inflation produced a primordial spectrum with oscillations, such that the 
CMB temperature power spectrum has no oscillations?  Can a $C_l$
spectrum with  
no oscillations really rule out inflation? In this paper we
investigate this possibility.

Suppose that the temperature spectrum measured by \pk\  turns out not 
to have oscillations. Note that as we see in Section \ref{sec.temp},
the new Boomerang \cite{b98} and Maxima \cite{maxima} data can be fit equally
well by spectra with and without oscillations, so we have no
indication either way as yet\footnote{
A different issue is whether the {\em single} peak that is clearly evident in
the current data is too sharp to be produced by defect models.
One of us answers this question in the affirmative in
\protect\cite{AAMoriond00}. 
}.
With future observations we can hope to decide whether or not the
temperature spectrum has oscillations.
What could we say about an inflationary model
that produced a temperature spectrum without oscillations, from our
other observations, those of the 
polarization and matter power spectra? We look at a model which
has oscillations in the primordial power spectrum producing a
temperature anisotropy spectrum with no oscillations, to see how this
affects the other spectra. We argue that such an
inflationary model is forced to have polarization and matter spectra
which are easily distinguishable from those of active models, at least
when the best data sets are finally in. Thus, even for the special
borderline cases we consider here the presence or absence of secondary
oscillations remains a robust test of the fundamental nature of the
primordial perturbations.


\section{Oscillations in present day power spectra}\label{sec.origin}

In this section we review the origin of the oscillations in the angular
spectrum for passive models, and discuss what factors can affect
them.  We develop our discussion based on the approximate
scheme of Hu and Sugiyama \cite{HS1,HS2} which is more intuitive and
gives a good enough account of the the physics for our immediate
purposes.  For the full calculations which are reported later in the
paper we use the full Boltzmann code CMBFAST\cite{cmbfast}.

The temperature anisotropy $\Delta_T \equiv \Delta T/T$ is expanded in
plane waves and Legendre polynomials \cite{MaBert,BE87},
\be
\Delta_T({\bf x},{\bf n}, \tau)&=&\int d^3{\bf k} 
e^{i{\bf k}\cdot{\bf x}}\Delta_T({\bf
k},{\bf n},\tau) \nonumber \\
&=&\int d^3{\bf k} e^{i{\bf k}\cdot{\bf x}}
\sum_l(2l+1)(-i)^l\Delta_{Tl}({\bf
k},\tau)P_l({\bf k}\cdot{\bf n}) . 
\ee
The Boltzmann equation must be solved for the photon phase space
distribution to get $\Delta_{Tl}({\bf k},\tau)$.

The observed sky is expanded in spherical harmonics $\Delta_T({\bf
x=0},{\bf n}, \tau_0)=\sum_{lm}a_{lm}Y_{lm}({\bf n})$, so to compare the
theoretical predictions to observations it is convenient to calculate
the angular power spectrum,
\be
C_{Tl}&\equiv&\langle |a_{lm}|^2\rangle  \nonumber \\
&=&(4\pi)^2\int k^2 dk P_i(k)|\Delta_{Tl}(k,\tau_0)|^2
\ee
where
$
<\Delta_{Tl}({\bf k},\tau)\Delta_{Tl}^*({\bf k'},\tau)>=\delta({\bf k}
-{\bf k'})P_i(k)|\Delta_{Tl}(k,\tau)|^2
$, $P_i(k)$ is the primordial spectrum, 
$\Delta_{Tl}(k,\tau)$ is the photon transfer function and $\tau_0$ is
the conformal time today. Homogeneity and
isotropy mean that these quantities depend only on the modulus of the
wave vector $\bf k$. It is important to note that the photon transfer
function does not depend on the initial power spectrum.

The angular anisotropy spectra for polarization and the
cross-correlation between temperature and polarization are defined
similarly,
\be
C_{El} &=& (4\pi)^2\frac{(l-2)!}{(l+2)!}\int k^2 dk P_i(k)|\Delta_{El}
(k,\tau_0)|^2,
\nonumber \\
C_{Cl} &=& (4\pi)^2\sqrt{\frac{(l-2)!}{(l+2)!}}\int k^2 dk P_i(k)
\Delta_{Tl}(k,\tau_0)\Delta_{El}(k,\tau_0).
\ee

The $\Delta_{El}(k,\tau)$ are rotationally invariant quantities 
related to the
$\Delta_{Pl}(k,\tau)$ that appears in the Boltzmann equation by a spin
lowering operator $\Delta_{El}(k,\tau)= -\eth^2 \Delta_{Pl}(k,\tau)$ 
\cite{zaldpol}.

The Boltzmann equations in Fourier space for the gauge-invariant scalar temperature 
and polarization
anisotropies are \cite{HS1,BE84,MaBert}
\be
\dot{\Delta}_T + (ik\mu + \dot{\kappa})\Delta_T &=&
-\dot{\Phi}-ik\mu\Psi 
+ \dot{\kappa}(\Delta_{T0}+i\mu V_b + \frac{1}{2}P_2(\mu)\Pi), \nonumber \\
\dot{\Delta}_P + ( ik\mu + \dot{\kappa})\Delta_P 
&=& \dot{\kappa}
\left( \frac{1}{2}(1-P_2(\mu))\Pi\right) .
\ee
where $\mu = {\bf k}\cdot{\bf n}$, $\kappa$ is the optical depth,
$\Psi$ and $\Phi$ are gravitational potentials, $V_b$ is the baryon
velocity and $\Pi = \Delta_{T2} +\Delta_{P0} +\Delta_{P2}$. Since
there is much less power in the polarization anisotropy than in the
temperature, $\Pi \approx \Delta_{T2}$, and can be ignored in the
temperature equation as the temperature monopole and baryon velocity
dominate during scattering. 

In order to understand the behaviour of the anisotropies, the
visibility function $\dot{\kappa} e^{-\kappa}$ can be approximated by
a delta-function at the time of recombination $\tau_*$. 
Solving the Boltzmann equation
gives us \cite{HS1}
\be
\Delta_{Tl}(k,\tau_0) &\approx& 
(\Delta_{T0}+\Psi)(k,\tau_*)(2l+1)j_l(k(\tau_0-\tau_*)) \nonumber \\
& & \nonumber \\
&+&
\Delta_{T1}(k,\tau_*)(lj_{l-1}(k(\tau_0-\tau_*)) -
(l+1)j_{l+1}(k(\tau_0-\tau_*))) \nonumber \\
& & \nonumber \\
&+&
(2l+1)\int_{\tau_*}^{\tau_0}d\tau(\dot{\Psi}-\dot{\Phi})
j_l(k(\tau_0-\tau))
\label{equ:boltzsolve1}
\\
\Delta_{El}(k,\tau_0) &\approx&
\frac{3}{4}(2l+1)\Delta_{T2}(k,\tau_*)\frac{j_l(k(\tau_0-\tau_*))}
{(k(\tau_0-\tau_*))^2}
\label{equ:boltzsolve2}
\ee

There are three terms that come from the surface of last
scattering. The first two terms contributing to the temperature
perturbation are the first two multipoles $\Delta_{T0}+\Psi$ and 
$\Delta_{T1}$, which come from
the intrinsic density fluctuations and velocity fluctuations on the
last scattering surface respectively. These are created during the
tight coupling epoch before last scattering, where the photons and
baryons are coupled via Compton scattering. The Bessel functions act
as projectors from fluctuations in $k$ on the last scattering surface
to fluctuations in $l$ that we see today. The quadrupole $\Delta_{T2}$
which produces photon polarization is created during
last scattering, as the electrons and protons recombine. 

The third term contributing to the temperature fluctuation is the
integrated Sachs-Wolfe effect (ISW), where photons travelling towards 
us are
perturbed by fluctuations in the gravitational field.
Large angles are dominated by the ISW effect, small angles by the
contributions made during scattering. Oscillations in the power
spectrum are created on small scales (inside the horizon at last
scattering) as shown below, so the ISW effect will not be relevant here.


With adiabatic initial conditions, on small scales during tight
coupling the monopole and
dipole terms are \cite{HS1}
\be
\Delta_{T0}(k,\tau) &\approx& (1+R)^{-1/4}\Delta_{T0}(k,0)\cos(kr_s(\tau)),
\nonumber \\
\Delta_{T1}(k,\tau) &\approx& -\sqrt{3}(1+R)^{-3/4}\Delta_{T0}(k,0)
\sin(kr_s(\tau)),
\ee
where $R$ is the ratio of baryons to photons normalised to $3/4$ at
equality and $r_s$ is the sound horizon. 
We can set $\Delta_{T0}(k,0)$ to $1$ for all $k$ as the initial 
conditions
are specified in the initial power spectrum.
The dipole (velocity contribution) oscillates out of phase with the
monopole (density contribution), and with reduced  
amplitude. So the temperature anisotropy today,
$\Delta_{Tl}(k,\tau_0)$, is dominated by the density fluctuation at
last scattering. 


From the solution to the Boltzmann equation
(\ref{equ:boltzsolve1}) we can see that at times shortly after $\tau_{*}$
the quadrupole behaves like
$
\Delta_{T2} \sim \Delta_{T1},
\label{equ:md}
$
so the behaviour of the polarization anisotropy today is dominated by
the velocity fluctuation during last scattering. 

So for adiabatic initial conditions the temperature perturbations on the 
surface of last scattering are
dominated by a cosine oscillation, and the 
polarization perturbations are
dominated by a sine oscillation.
The bessel functions act as projectors from Fourier mode $k$ to
multipole $l$, so for large
$l$ we only need to consider large $k$ contributions to the angular
power spectra to see their oscillatory structure,
\be
C_{Tl} &\sim& P_i(l/(\tau_0-\tau_*))
\cos^2(lr_s(\tau_*)/(\tau_0-\tau*)), \nonumber \\
C_{El} &\sim& P_i(l/(\tau_0-\tau_*))
\sin^2(lr_s(\tau_*)/(\tau_0-\tau_*)), \nonumber \\
C_{TEl} &\sim& P_i(l/(\tau_0-\tau_*)) \sin(2lr_s(\tau_*)/(\tau_0-\tau_*)),
\ee
for adiabatic initial conditions.

The oscillatory contributions to the $C_l$s come only from the physics
of the photon-baryon fluid, which is the same regardless of initial
power spectrum. Therefore we may try to produce a
non-oscillatory $C_{Tl}$ spectrum with an oscillating primordial
spectrum. We will show this in the next section.

We see that $C_{Tl},C_{El}$ and $C_{TEl}$ all oscillate with the same
frequency, but $C_{Tl},C_{El}$ are half a period out of phase and
$C_{TEl}$ is a quarter of a period out of phase with both $C_{Tl}$ and
$C_{El}$.
Therefore we cannot erase oscillations in more than one of these
spectra at a time simply by changing the initial power spectrum of
perturbations. If we produce a non-oscillatory spectrum of temperature
perturbations, we must necessarily produce enhanced oscillations in
the polarization and temperature polarization cross-correlation spectra.
The question arises, however, of whether these enhanced oscillations
would be observable. We look at this in section \ref{sec.polarization}.

The effect on the matter power spectrum is much simpler. As we look at
this in Fourier space and we are only considering linear behaviour,
the Fourier modes do not couple and we can relate the spectrum today
directly to the primordial spectrum with a transfer function:
\be
P_f(k)=T^2(k)P_i(k),
\ee
where $P_f(k)=|\delta_f(k)|^2$ is the power spectrum today.

Transfer functions for all species fall off for $k>k_{eq}$ as
perturbations inside the horizon decay during the radiation
dominated era.
The baryon
transfer function has oscillations inside the horizon before and
during 
recombination,
due to the oscillations in the photon-baryon fluid as described
above, while the CDM transfer function is smooth since CDM interacts
only gravitationally. After recombination however, the baryons
stop interacting with the photons, effectively have no pressure, and
only interact gravitationally, falling into the gravitational wells
created by the CDM. The CDM
and baryons therefore end up with the same transfer function today. 
In models
with low baryon density the transfer function is smooth today, for
models with high baryon density the oscillations may not have
disappeared by the present day.
In general, there could also be a hot or warm component to the dark
matter, and these would also tend to damp the oscillations.
For models in which the present day transfer function has no
oscillations, oscillations in the primordial spectrum would still be
present today. If the transfer function still has oscillations, eg due 
to a large baryon fraction, we
would expect them to be at least partially cancelled by the primordial
oscillations. In Section \ref{sec.matter} we calculate the present day
spectrum and discuss whether these oscillations are observable.

We note that while we focus on the standard adiabatic picture here,
one could construct a very similar discussion for the isocurvature
case. Although specific aspects of the oscillations are different
for isocurvature perturbations, the tendency for the fluid to
oscillate and the resulting impact on the different observations is
similar.  In fact, a more general analysis would include a combination
of all possible perturbation modes, as was done recently in
\cite{BMT00}. It is  clear from the results in \cite{BMT00} that the
differentiating power of the polarization is present even when general
mixtures of perturbation modes are considered.  Introducing a wider
fitting parameter space will naturally require more precise data to
pin down the parameters, but we expect the main thrust of our analysis to
persist under a more general multi-mode treatment.


\section{Initial Power Spectrum}\label{sec.init}

There has been much interest recently in models of inflation which
produce a broken-scale-invariant spectrum of perturbations. There is a
model which produces oscillations in the primordial spectrum from an
oscillating inflaton field \cite{muk,lidmaz}, and there has been work
on models with two fields in a potential with a step, producing
oscillations in the spectrum below some scale \cite{starob,lesgourgues}. 
Also multiple
periods of inflation could produce oscillations in the spectrum
\cite{sarkar}. 

We have chosen not to use a specific model to produce
the initial spectrum, but to use a spectrum which completely cancels 
oscillations in the temperature spectrum
created during the tight coupling era, as a worst case scenario, as
this would be most confused with active models.
The two models in this paper are examples for illustration of the
ideas and analysis of how well they fit the data. They are
intended to demonstrate some general features of this type of model.
While Barrow and Liddle \cite{B-n-L} argue that it might not be
possible to construct a realistic inflaton that can produce the sort
of primordial spectrum we discuss here, they only consider
single-dimensional inflation space.  If the data really do not show
oscillations it is quite likely that some inflation enthusiast will
find a model that tries to evade the secondary oscillation test.   In
any case our discussion is valid for the more general question passive
vs. active perturbations.

\begin{figure}
\setlength{\unitlength}{1cm}
\centerline{\hbox{\psfig{file=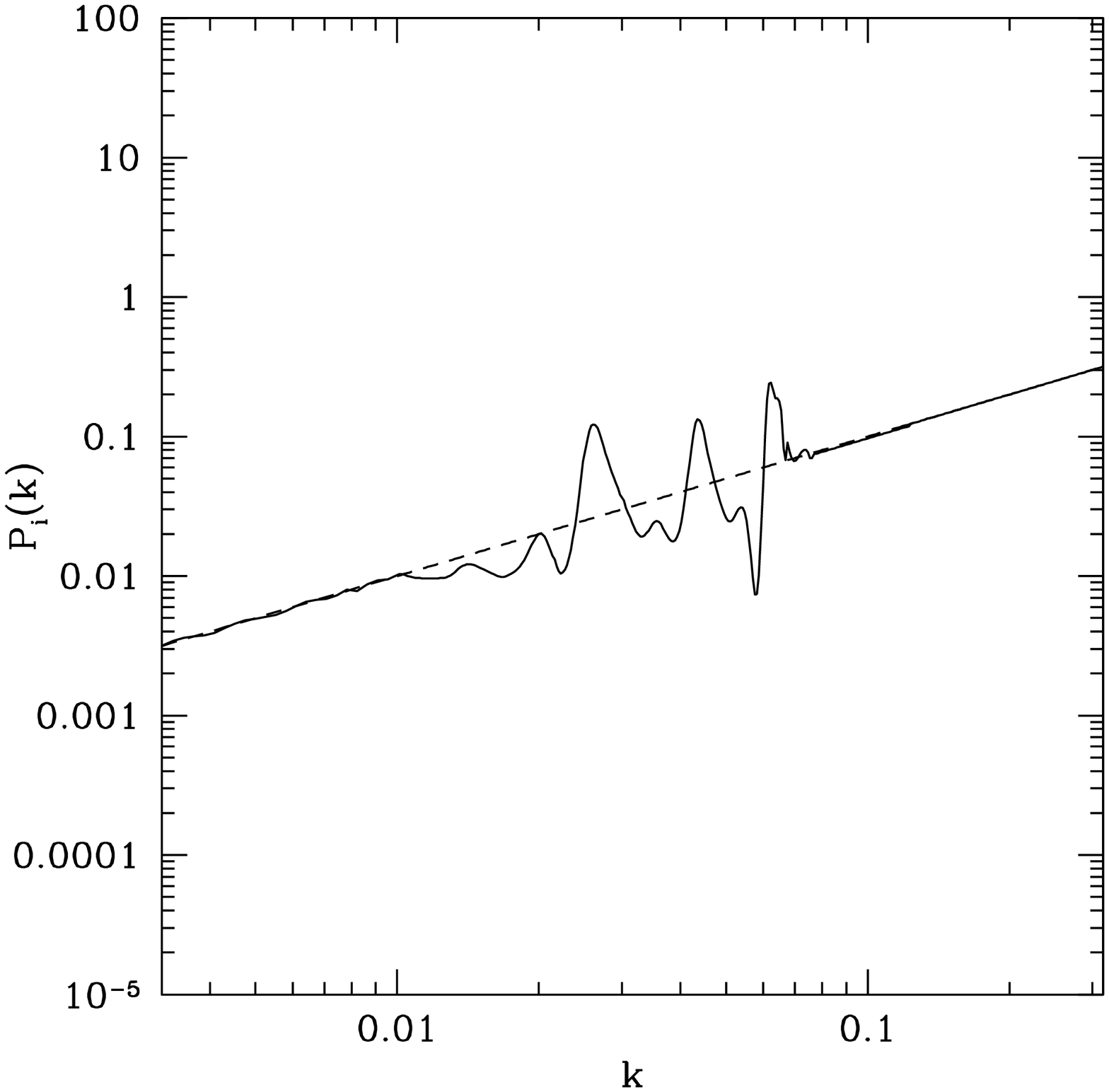,height=6cm,width=7cm}
\hspace{2cm}\psfig{file=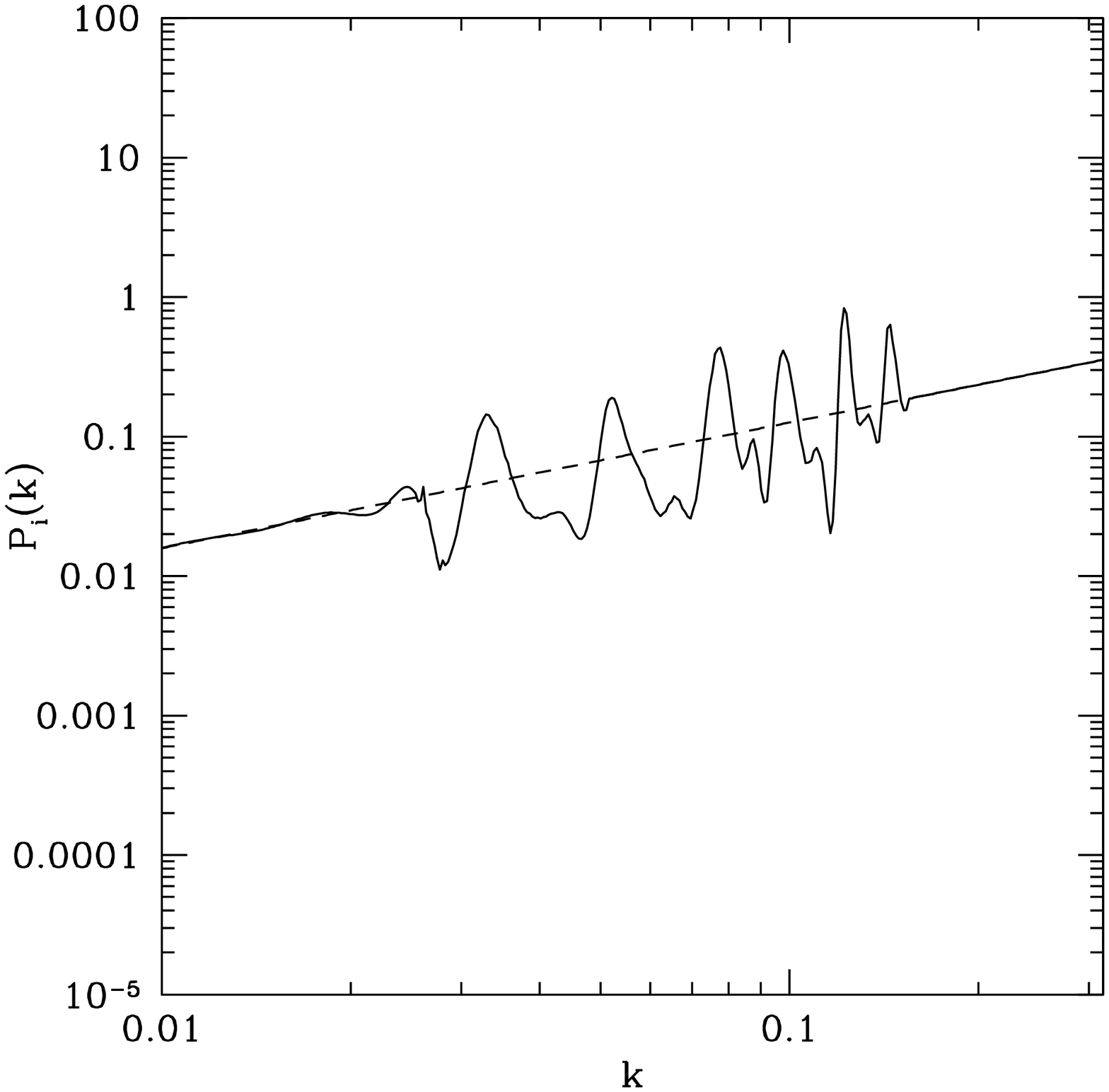,height=6cm,width=7cm}}}
\caption{Initial spectra with oscillations for \ma\ (left) and \mb\
(right). The dashed lines are the scale-invariant spectra.} 
\label{fig.initps}
\end{figure}

We have looked at two models: one with a cosmological constant and a
low value of $H_0$ (\ma) and one with a higher Hubble constant and a
tilted initial spectrum (\mb). \ma\ has $\Omega_b=0.061$,
$\Omega_{m}=0.32$, $\Omega_{\Lambda}=0.68$ and $h=0.47$. \mb\ has $\Omega_b=0.05$,
$\Omega_{m}=0.3$, $\Omega_{\Lambda}=0.7$ and $h=0.7$.
Figure \ref{fig.initps} shows the initial spectra with
oscillations for the two models. Also shown are non-oscillatory 
spectra for comparison.  

\section{Temperature Anisotropy Spectrum}\label{sec.temp}

\begin{figure}
\setlength{\unitlength}{1cm}
\centerline{\hbox{\psfig{file=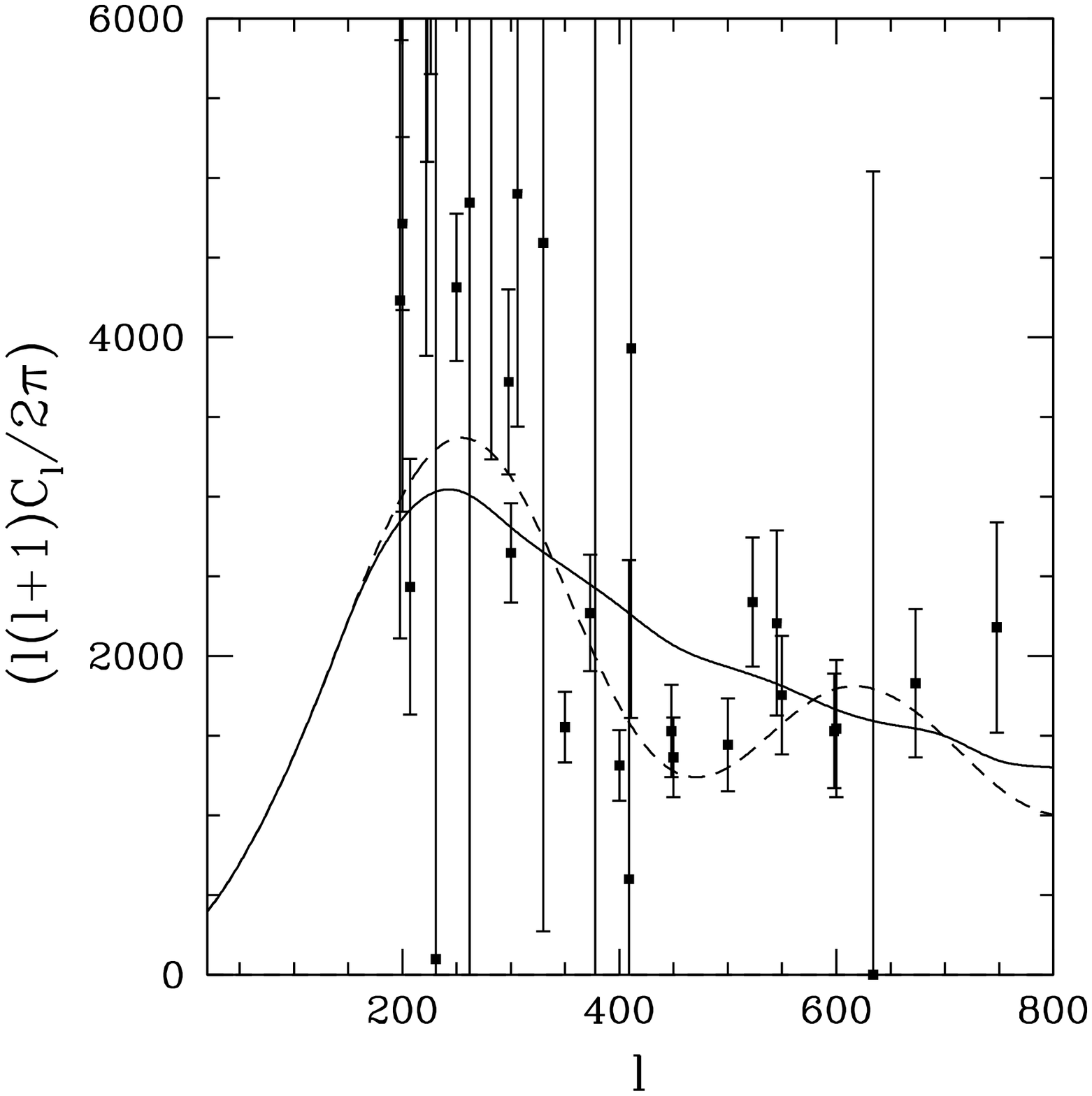,height=6cm,width=7cm}
\hspace{2cm}\psfig{file=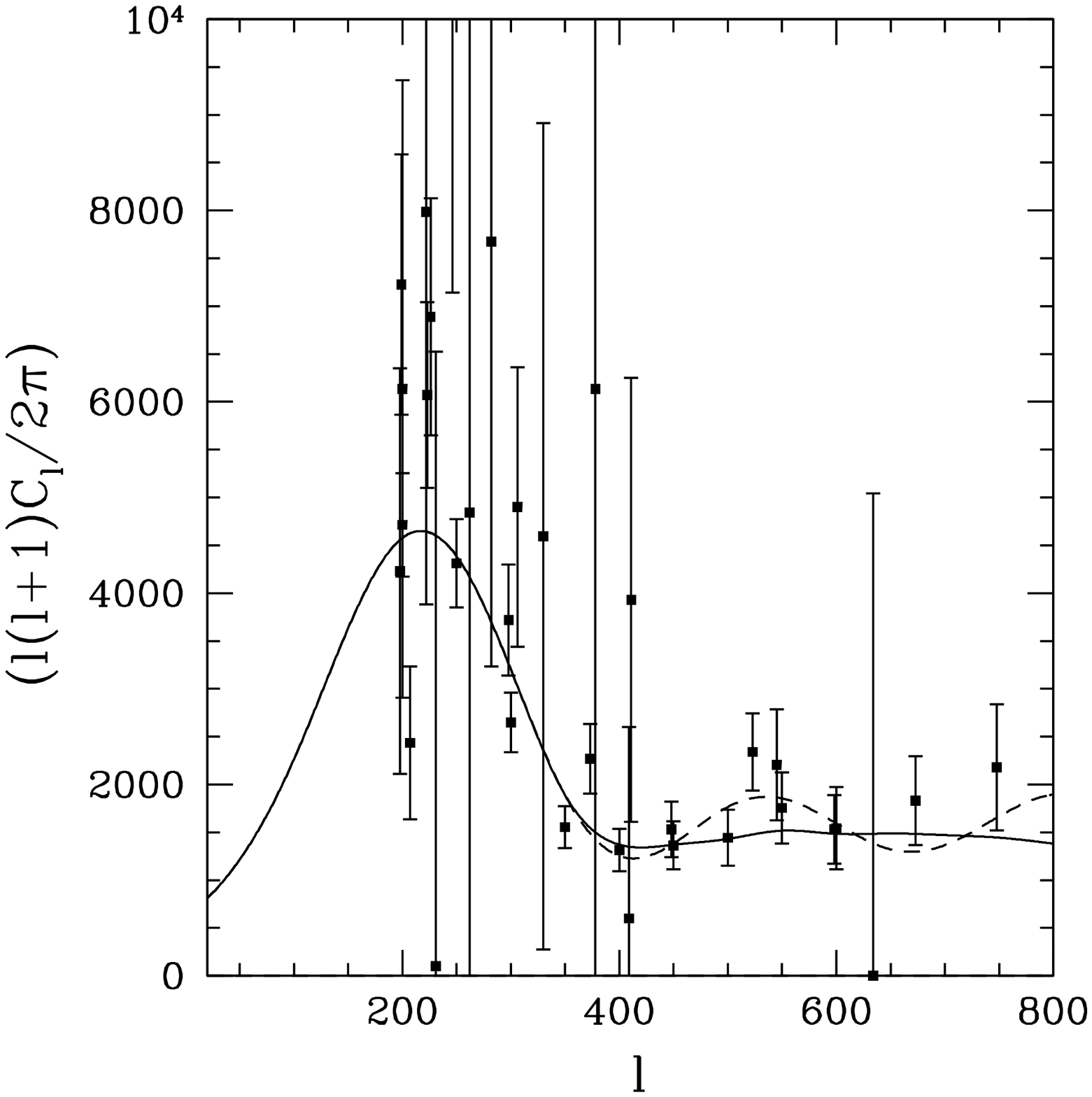,height=6cm,width=7cm}}}
\caption{Temperature anisotropy spectra for models with and without 
oscillations shown with current data 
with effective $l$ of 180 or higher. Left: \ma, right:
\mb} 
\label{fig.currentcl}
\end{figure}

First we will compare the models to existing data on the microwave
background. We have chosen to look only at high $l$ since the spectra
agree at low $l$. 
Figure \ref{fig.currentcl} shows the usual and modified 
$C_{Tl}$ spectra for the two models, with all the data points
currently available with an effective $l$ of over 180. This includes
Saskatoon \cite{sask}, MSAM \cite{msam}, Python
\cite{python,web-coble}, MAT \cite{mat}, CAT \cite{cat}, OVRO
\cite{ovro}, SUZIE \cite{suzie}, Boomerang \cite{b98} and MAXIMA
\cite{maxima} data.


In order to compare the predictions from the models to experimental
data we use the package Radpack \cite{web-knox}. This calculates the \ch\
between data and model in band powers, using a transformation of
the $C_l$s into approximately Gaussian distributed quantities \cite{bjk2}.

\ma\ with a scale invariant initial spectrum has a \ch\ per degree of freedom
of 2.9 and the same model with an oscillating spectrum has \ch\ per
degree of freedom of 3.8. \mb\ has \ch\ per degree of freedom of 0.94
with a non-oscillatory initial spectrum and 0.95 with an oscillating
primordial spectrum, so fits the data very well. 
Introducing oscillations into the primordial power spectrum does not
alter the level of fit appreciably. 
Clearly we cannot say anything yet 
about the presence or absence of oscillations.

\begin{figure}
\setlength{\unitlength}{1cm}
\centerline{\hbox{\psfig{file=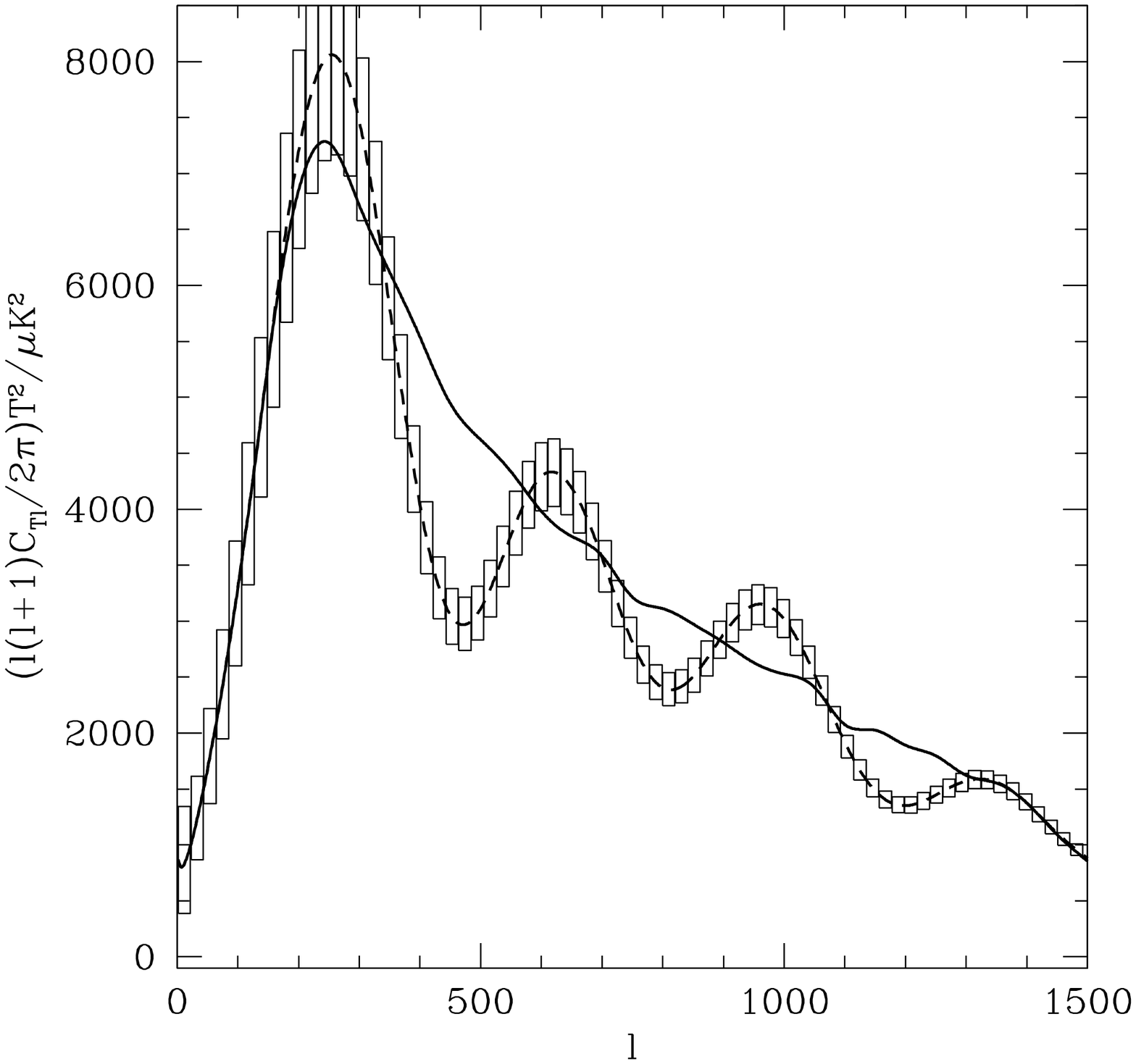,height=6cm,width=7cm}
\hspace{2cm}\psfig{file=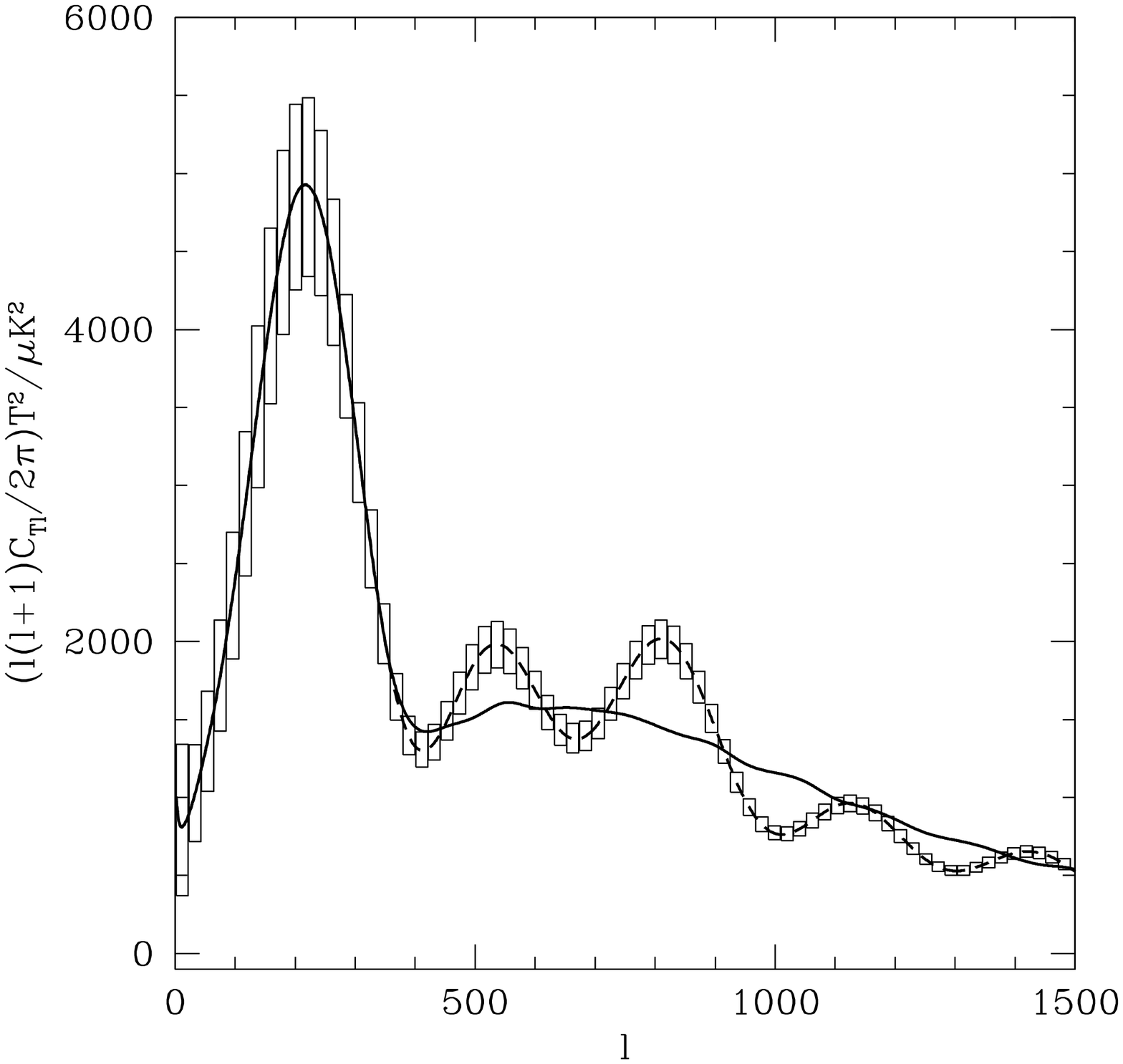,height=6cm,width=7cm}}}
\caption{Temperature anisotropy spectra for models with and without 
oscillations shown with predicted
errorbars as measured by \pk. Left: \ma, right: \mb} 
\label{fig.futurecl}
\end{figure}

Future observations should be able to tell if there are oscillations
in the CMB temperature spectrum. 
Figure \ref{fig.futurecl} 
shows errorbars (including the beam and uniform, uncorrelated noise,
assuming foregrounds have been removed) predicted for the \pk\  
satellite for $C_{Tl}$ spectra
with non-oscillatory spectra. Also shown are the spectra produced
by an oscillatory primordial spectrum. 

The errorbars are calculated in the standard way, 
assuming that the $a_{lm}$ are Gaussian
distributed and that there is spatially-uniform, Gaussian noise 
\cite{knox95,bet}. 
The $1 \sigma$ error bar in $C_{Tl}$ is
$
\Delta C_{Tl}  = \sqrt{2/(2l+1)f_{sky}} (C_{Tl} +
\omega_T^{-1}B_l^{-2}) 
$
where $ \omega_T B_l^2 = \sum_c
\omega_{T(c)}B_{l(c)}^2 $, $\omega_{T(c)}^{-1}$ is the
variance of the noise in the temperature measurement in channel c, and
$B_{l(c)} \approx \exp{(-l^2\sigma_{(c)}^2/2)}$ is the Legendre transform 
of the
beam for channel c. The angular resolution of the beam is given by
$\sigma_{(c)} = \theta_{fwhm(c)}/\sqrt{8\ln 2}$ where $\theta_{fwhm}$ 
is the
full width at half max of the beam in radians.
Incomplete sky coverage is partially taken into account by the factor 
$f_{sky}$
which is the fraction of sky used to estimate the $C_{Tl}$.
For \pk\ we have used $\omega_T^{-1}=(0.011\mu K)^2$, combining the 
143 GHz and
217 GHz channels at full width half max of $8.0^{'}$ and $5.5^{'}$
respectively, and $f_{sky}=0.33$ \cite{planck}. For the MAP satellite
\cite{map} we used $\omega_T^{-1}=(0.106\mu K)^2$, with the highest
frequency channels (40 GHz, 60 GHz and 90 GHz) at full width half max
of $0.47^\circ$, $0.35^\circ$ and $0.21^\circ$ respectively and
$f_{sky}=0.66$.
The estimates of the errors are
averaged in arbitrary bins to make the plots clearer.

In order to determine whether or not future experiments will be able
to 
distinguish 
between two models we need a different measure from the \ch\ statistic 
used to compare theory 
with observation. Since now we have two models which we wish to
compare on an equal footing, we need a statistic that is symmetric in
the two theories. We consider simulating $\hat{C_l}$s from each theory 
 with mean \cl\ and standard deviation $\Delta C_l$. We use a
Normal distribution  to approximate the distribution of the \cl
s since we are only interested in the higher $l$s. 

There is a standard way to test if two samples have the same
underlying mean: if $\widehat{C_l}^1$ and $\widehat{C_l}^2$ are both
Gaussian-distributed with the same mean but possibly different
variances then their difference will be Gaussian-distributed with mean
zero and variance $(\Delta C_l^1)^2 + (\Delta C_l^2)^2$. So for each
$l$ 
we
should use the statistic
\be
\chi^2(l) = \frac{(\widehat{C_l}^1 - \widehat{C_l}^2)^2}{(\Delta
C_l^1)^2 
+ (\Delta
C_l^2)^2}
\ee
and sum them over $l$.

In fact we can use this idea to give a measure of how likely it is to
confuse two
models without performing simulations. The expectation of $\chi^2(l)$
is 
\be
d_l = 1 + \frac{(C_l^1 - C_l^2)^2}{(\Delta C_l^1)^2 + (\Delta
C_l^2)^2}.
\ee
We will interpret the sum over the $d_l$ in terms of a \ch\
distribution, in
order to find a probability with which the two theories could be
confused. 
We define $d$ to be the average of the $d_l$.
Note that if the two theories are exactly the same, $d=1$ as
expected. We can use this measure to see how similar two theories are
ideally, that is in the limit of no instrumental effects. To look at
how well a particular experiment will distinguish the theories we put
the noise, beam and fractional sky coverage effects into the spread in
the \cl s, the $\Delta C_l$ as described above.

Comparing the \ma\ $C_l$ spectra with and without oscillations, using the MAP
noise and beam parameters, gives a value for $d$ of 2.6, which if
interpreted as a \ch\ per degree of freedom would mean that confusing
the two models is ruled out at a $\sigma$ level of 43. For \mb\ $d$ is
1.2, which corresponds to a $\sigma$ level of 3.9. The \mb\ spectra are
harder to distinguish, since having a tilted initial spectrum suppresses the
oscillations in the $C_l$s. 
The \pk\
satellite will be more discriminating: with \pk\ characteristics
$d=4.8$ for \ma\ and 5.0 for \mb.

Attempts have been made \cite{bouchet-fore,knox-fore,teg-fore} to
estimate the effect of foregrounds on our ability to measure the
angular power spectrum. 
We can use the degradation factors for the temperature error bars in
\cite{teg-fore} to calculate $d$ for the satellites taking into
account their foreground models. The degradation factors $D_l$
increase the errorbars,  $\Delta C_l^{FG} = D_l \Delta C_l$. 
We hope that the estimates for $D_l$ given in \cite{teg-fore} would
not vary too much with different underlying cosmological
models. We use an upper limit $D$ for the $D_l$ to give some room for
variation. 

This gives us
\be
d^{FG} = 1+(d-1)/D^2
\ee
For the main foreground scenario in
\cite{teg-fore} $D_l \leq 1.01$ for all $l$ values and both satellites which 
means that $d$
is essentially unchanged. For their more pessimistic model $D_l \leq 2$
which leads to $d^{FG} \geq 2.0$ for \pk\ for both models, and 
$d^{FG} \geq 1.4$ and
$d^{FG} \geq 1.0$ for MAP for \ma\ and \mb\ respectively. 
MAP will not be able to tell whether or not a tilted model has
oscillations, but \pk\ can distinguish in both cases we consider here.

To get the above values for $d_l$, we have not used $l$ values below
50, as $C_l$ are not Gaussian distributed for low $l$. The models
we are interested in do not differ at those $l$, so $d$
would decrease if we included them, but not enough to make a
difference to the conclusions we draw. We have calculated each $d_l$
separately, without binning, but this would make errorbars smaller and
therefore $d$ even bigger.


\section{Matter Power Spectrum}\label{sec.matter}

In this section we look at how the altered primordial spectrum affects
the present day matter power spectrum and whether these changes are
observable. We would like to compare the model
predictions for the matter power spectrum with current data, to see
what we can already say about models with oscillations in the
primordial spectrum.
In the currently favoured models the transfer functions are smooth, and
any features in the initial spectrum would show up in the
present day matter power spectrum.  For such models
the power spectrum of the matter density field is the best
probe of possible oscillations in the initial spectrum.


\begin{figure}
\setlength{\unitlength}{1cm}
\centerline{\hbox{\psfig{file=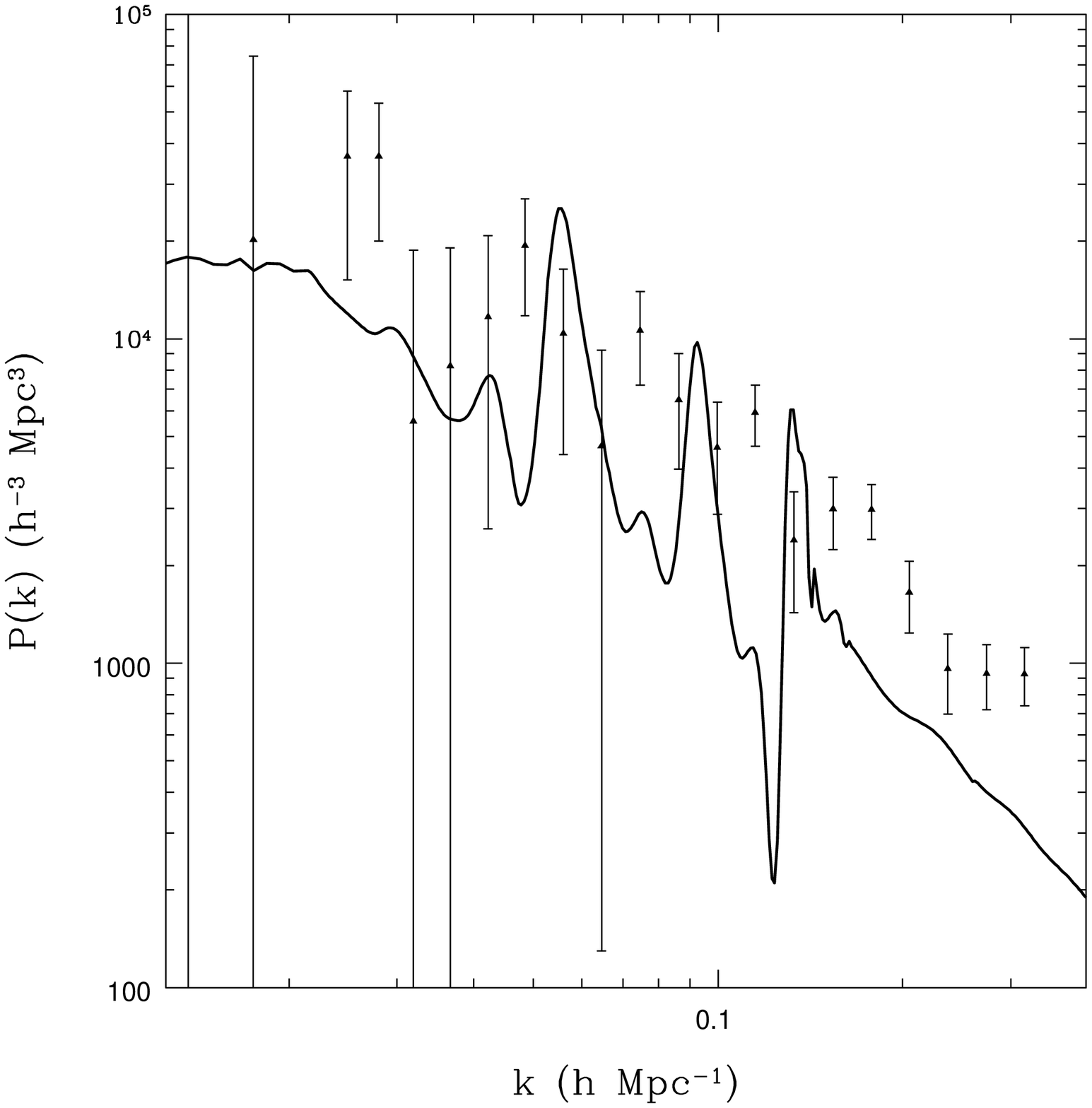,height=6cm,width=7cm}
\hspace{2cm}\psfig{file=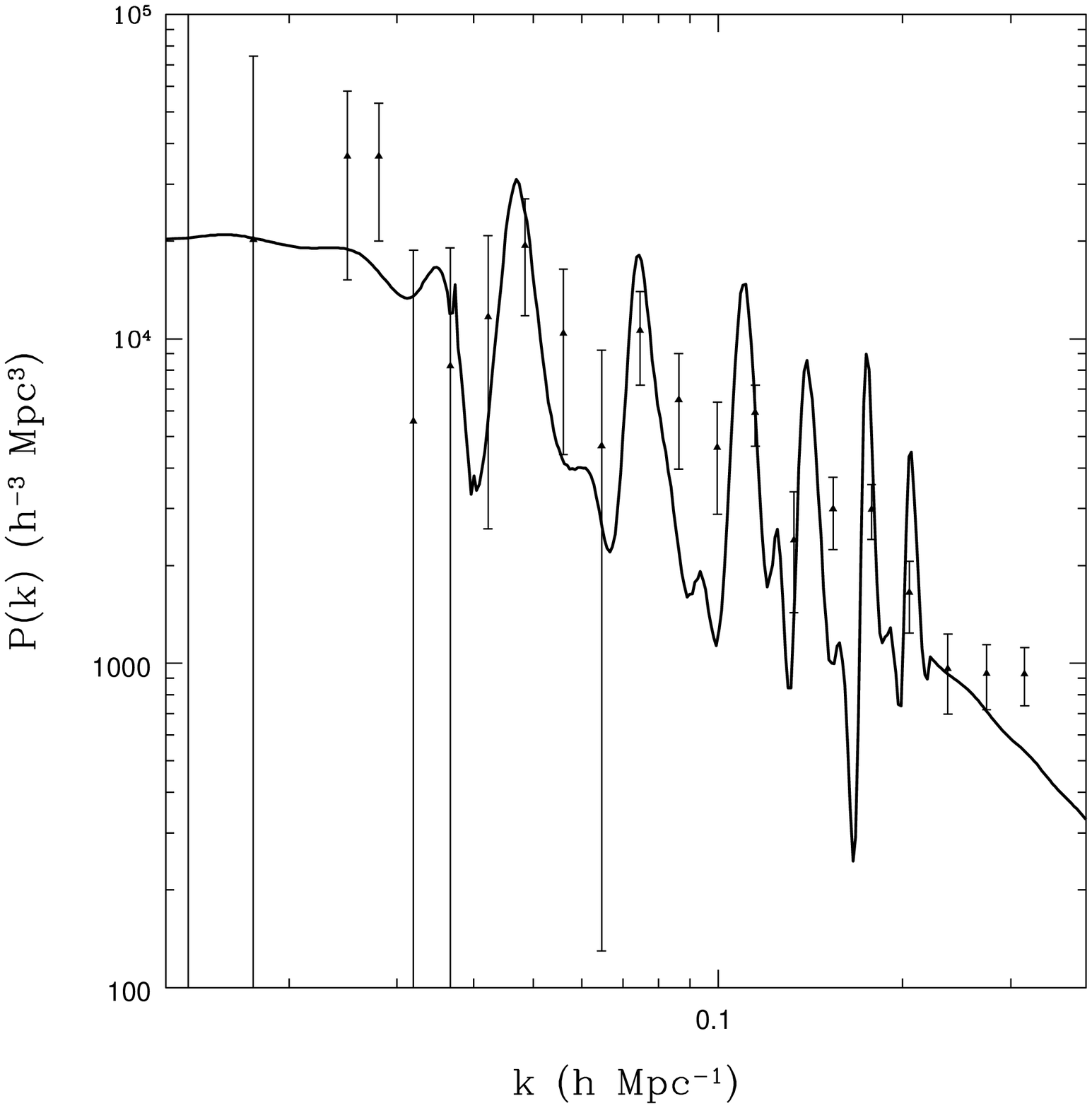,height=6cm,width=7cm}}}
\caption{Present day matter power spectrum from the models with 
oscillations in the primordial spectrum, shown with data from
\protect\cite{hamilton}.  Left: \ma, right: \mb.}
\label{fig.currentmat}
\end{figure}

Figure \ref{fig.currentmat} shows the 
matter power
spectrum from the models with and without oscillations and data 
from \cite{hamilton}. 
We use the decorrelated linear spectrum which has effectively
uncorrelated points. It is important to have independent points since
oscillations would be smoothed out in correlated data. 
We have calculated
the \ch\ for our models binned in the same way as the data. 
For \ma\ 
with a scale invariant initial spectrum the \ch\ per degree of freedom
is 3.2 and with oscillations in the initial power spectrum it is
3.4. 
For \mb\ with oscillations the \ch\ per degree of freedom is 1.2 with
and without oscillations. This model fits the data well but we cannot
say whether or not the spectrum has oscillations.


We can look at averaged values, such as $\sigma_8$. For \ma\ this is
0.53 and 0.56 without and with oscillations in the
primordial spectrum, normalised to COBE. For the \mb\ the values are
0.68 and 0.74 without and with oscillations. 
We see that oscillations in the spectrum have very little effect on
the value of $\sigma_8$. We would expect this, as we have not changed
the overall tilt of the spectrum.


 It does not seem that the current data on the matter 
power spectrum can tell us much about the presence or absence of 
oscillations as we have discussed here. In order to draw conclusions
about this we need data points with a smaller coherence length. We 
hope that new data from
redshift surveys will constrain the power spectrum much more than has
been possible so far \cite{vogeley}.

One possible complication would be if the preferred
model turns out to have an oscillatory matter transfer function (eg
because the matter has a high proportion of baryons).  Even in that
case, with good enough data it might be possible to tease apart the
different effects.


\section{Polarization in the microwave background}\label{sec.polarization}

As we saw in Section \ref{sec.origin}, the power in the polarization
anisotropy oscillates out of phase with that of the temperature
anisotropy, so the extra power on certain scales in the initial
spectrum leads to enhanced polarization oscillations. Here we show
examples for our models.  Note that active models have completely
different polarization spectra so there is no opportunity for
confusion \cite{seljak-2,spergel}.

\begin{figure}
\setlength{\unitlength}{1cm}
\centerline{\hbox{\psfig{file=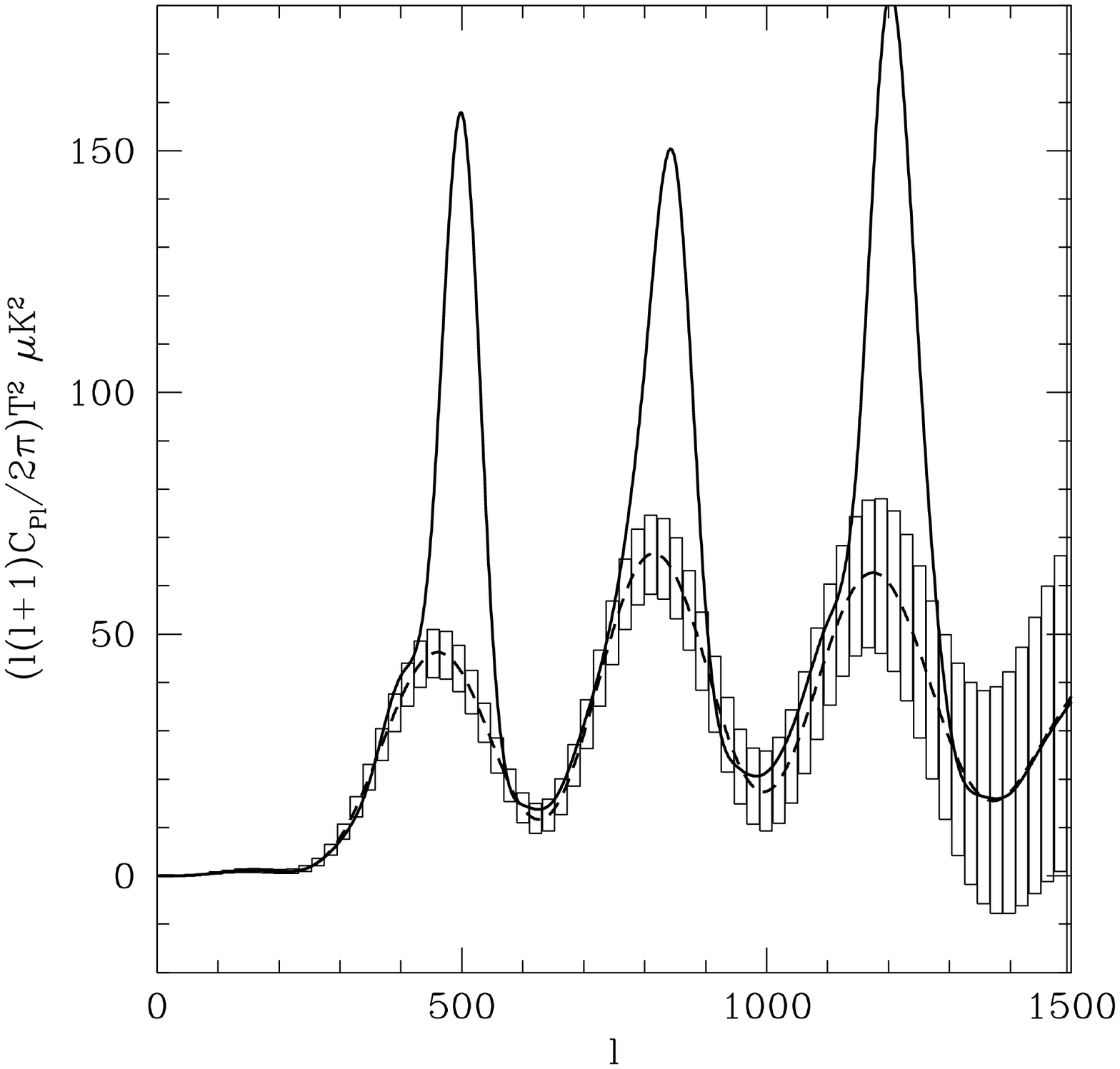,height=6cm,width=7cm}
\hspace{2cm}\psfig{file=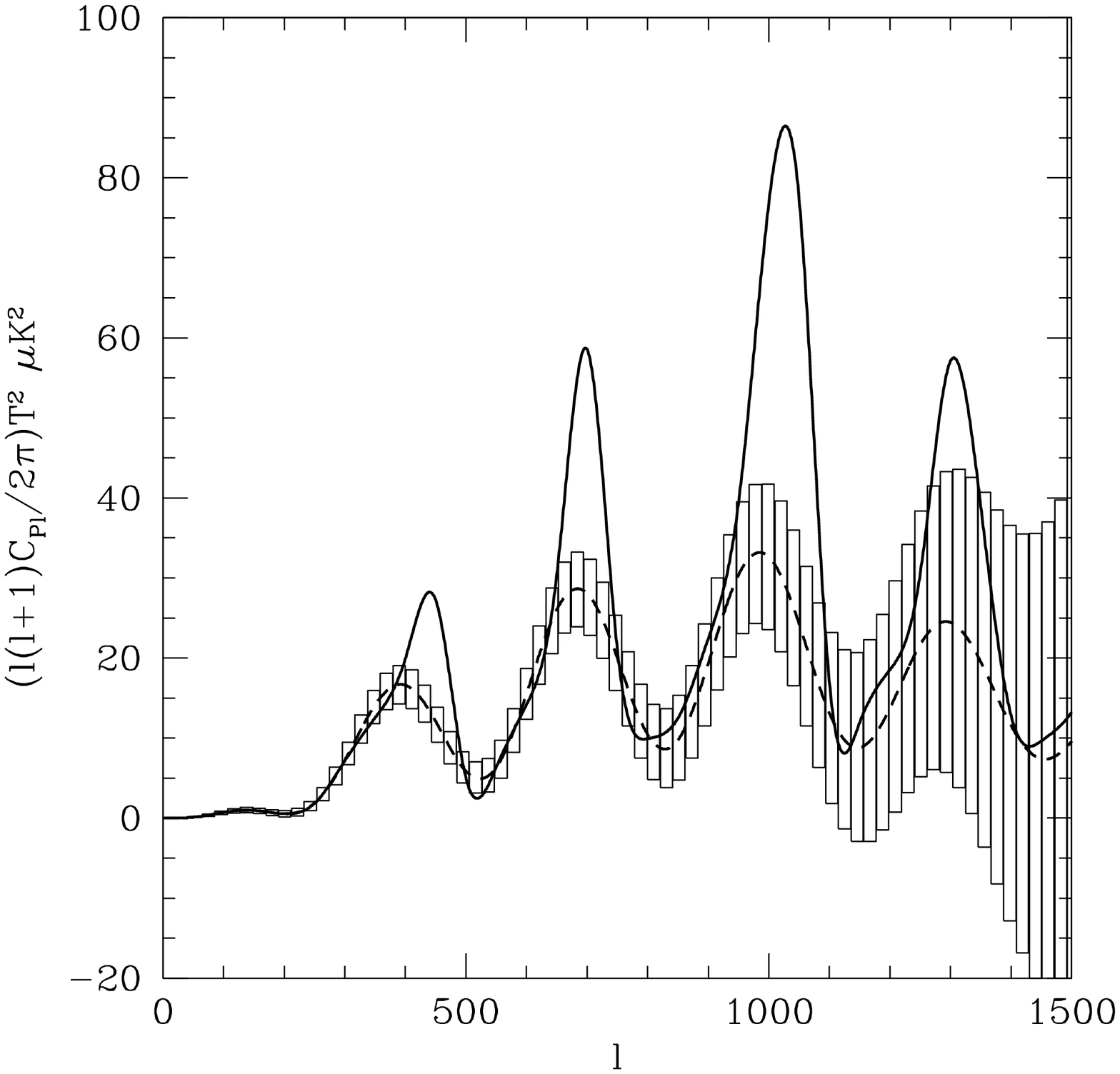,height=6cm,width=7cm}}}
\caption{Angular polarization power spectra for models with and
without oscillations, with errorbars forecast for the \pk\
satellite. The errorbars are shown for the models with scale-invariant
initial spectra. Left: \ma, right: \mb} 
\label{fig.futurepl}
\end{figure}

Figure \ref{fig.futurepl} shows the angular power 
spectra for polarization for the
two models using a scale invariant primordial spectrum with 
$1 \sigma$ errors \cite{zaldpol} expected from the \pk\ 
satellite. The parameters used are the same as for the temperature
spectra earlier.

The figures show clearly the enhancement of the polarization
spectrum. To see if these differences are observable we have used the
same measure $d$ as in Section \ref{sec.temp} to compare two models. 
Using errorbars predicted for \pk\, $d=6.5$ for \ma\ and 2.8 for
\mb. For MAP the values are 1.01 and 1.00 respectively. 
So we would not consider these two curves different
under MAP observations, but \pk\ should be able to distinguish
them.

In the main (supposedly most realistic) model of polarization
foregrounds from \cite{teg-fore} the degradation factor $D_l$ for errorbars
on the angular polarization spectrum is less than 2 for \pk\ at all $l$, which
gives $d^{FG} \geq 2.4$ for \ma\ and 
$d^{FG} \geq 1.5$ for \mb. So \pk\ will be able to tell
these models apart even in the presence of this level of foregrounds. The more
pessimistic model gives $D_l < 10$ for \pk\ at $l > 50$, giving
$d^{FG} > 1.06$ for \ma\ and $d^{FG} > 1.02$ for \mb, 
so if the foregrounds are this bad even \pk\ will not
separate these spectra.

\begin{figure}
\setlength{\unitlength}{1cm}
\centerline{\hbox{\psfig{file=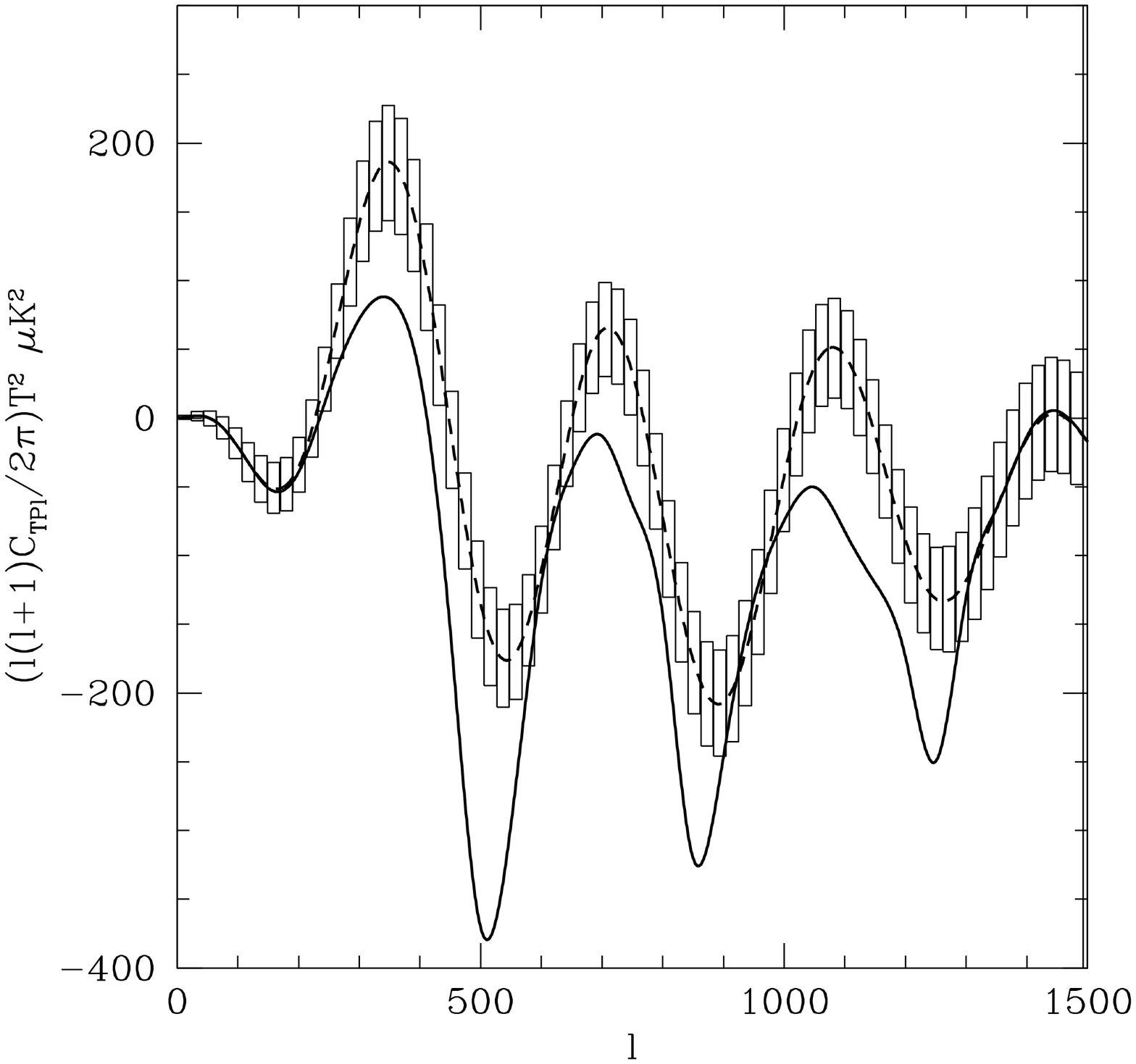,height=6cm,width=7cm}
\hspace{2cm}\psfig{file=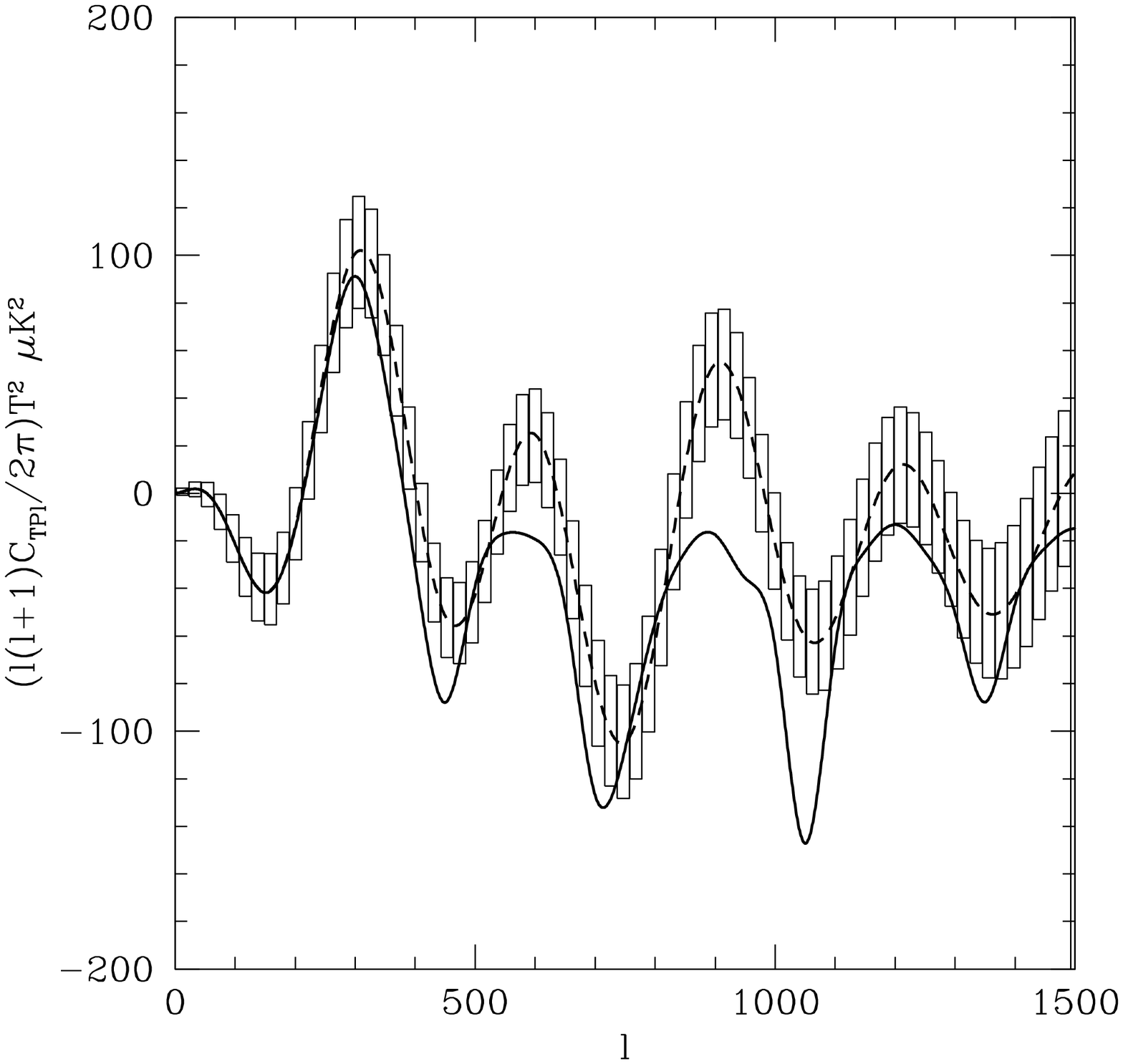,height=6cm,width=7cm}}}
\caption{Temperature-polarization cross-correlation for models with and
without oscillations, with errorbars forecast for the \pk\
satellite. The errorbars are shown for the models with scale-invariant
initial spectra. Left: \ma, right: \mb} 
\label{fig.futurexx}
\end{figure}

Figure \ref{fig.futurexx} shows the
temperature-polarization cross-correlation for the various models.
Here again we can see the shift in oscillations produced by altering
the primordial spectrum. 
Looking at the cross-correlation power should be more valuable
than the polarization on its own, for two reasons. The first is that
the noise begins to
dominate at higher $l$ for temperature than for polarization, so we
will get a clearer signal for the cross-correlation than for the
polarization alone. 
Using errorbars predicted for \pk, $d=5.0$ for \ma\ and
3.2 for \mb. For MAP the values are 1.09 and 1.01
respectively. 
These are too low to allow MAP to distinguish between the models, but
\pk\ would not be confused.  
We see that for these models
the cross-correlation is in fact not such a good discriminator between
models with and without oscillations in the temperature spectrum. This
is because although the errorbars are smaller, the models are closer
together. The effect of oscillations in the primordial spectrum is
clearer in the present day
polarization spectrum than in the temperature-polarization
correlation since in the polarization the oscillations are completely
out of phase, but in the cross-correlation they are only half a period
out of phase.

There are no estimates of the degradation factor for the
cross-correlation in \cite{teg-fore}. We make a rough estimate using
the fractional errors on $C_l$s with foregrounds obtained by
\cite{bouchet-fore}. These give degradation factors slightly larger
than the main foreground scenarios in \cite{teg-fore} for the
temperature and polarization spectra as found be \pk. We find $d^{FG}
\gtrsim 1.6$ for \ma\ and  $d^{FG} \gtrsim 1.4$ for \mb, hence we
estimate that with this level of foreground contamination the \pk\
data should discriminate between these models with and without
oscillations.

The second reason for looking at the cross-correlation power is 
that a strong prediction for active
models is that due to their incoherence the cross-correlation in
models where active sources play the major role in structure formation
will be close to zero. We have calculated $d$ for the
cross-correlation using the 
models with oscillations in the primordial spectrum and a `model' where
all the $C_{TEl}$ are zero, to see if the cross-correlation could be
consistent with zero. The value of $d$ for \ma\
is 15.4, and for \mb\ the value is 13.5. We
would not consider either of these cross-correlation spectra to be
consistent with a vanishing spectrum. Therefore an observation of a
smooth temperature anisotropy spectrum and vanishing
cross-correlation would be strong evidence against passive fluctuations.


\section{Discussion}

We have investigated the secondary oscillation test, which is
recognized as the most fundamental way in which scenarios with
perturbations from inflation can be falsified. While current data are
not good enough for the test to be applied, we look toward the future
when much better data will be available. We have studied particular
situations where the inflationary model conspires to evade the test by
building oscillations into the inflaton potential to effectively
`cancel out' the natural oscillations that would appear in the CMB
anisotropies.  We have shown that such conspiratorial models will
produce the characteristic oscillations in other observable spectra,
and thus will be unable to evade the test.


One might think that reionization could smooth the peaks of the
angular spectra and confuse us, but reionization also produces a
different signal in the polarization.
Reionization damps the temperature perturbations on all scales smaller
than the sound horizon at the start of the
reionization. Homogeneous reionization causes approximately
exponential damping in $l$ and so cannot alter the oscillations in the
spectrum. In order to do this one would have to have damping starting
on smaller scales than the sound horizon at recombination; clearly
the reionization scale is larger than that of recombination. 
Reionization damps the polarization spectrum as well but also produces 
more polarization on large scales due to
rescattering at the second `last scattering surface', resulting in a
boost at low $l$ in the polarization angular spectrum.
The oscillations in the spectra might be washed out
enough to be unobservable if the optical depth to reionization is
large enough, but increasing optical depth increases the extra power
in the polarization at low $l$ \cite{zald-pol-reio}, providing a
strong indication of the presence of reionization.
So inflationary perturbations with homogeneous reionization
could not evade the secondary oscillation test. 

The situation might be different in non-standard reionization where
the visibility function is broader, but this causes a large amount of
damping and is probably ruled out already by the temperature
observations. 
In inhomogeneous reionization models the first order temperature
spectrum comes entirely from the 
homogeneous reionization background. The second order effect due to
the inhomogeneities is too small compared to the first order to make
any difference to the peaks \cite{inhom-reio-1,inhom-reio-2}.


Work has been done recently on models with both active and passive
fluctuations \cite{strings-n-inflation}. The application of the fine tuning we have done
here to those models represents an even more exotic corner of parameter
space which we have not examined.

The matter power spectrum is affected by a number of issues
(such as the baryon fraction) which make the interpretation of the
presence or absence of oscillations less than clear-cut.
However, we have shown that the polarization power spectrum and the
temperature-polarization cross-correlation will offer a much more
straightforward test, assuming the cosmic signal can be separated from
the foregrounds. It is these spectra we are relying on to draw our
optimistic conclusions.

There is no question that the models we do consider are viewed by most
cosmologists (including ourselves) as extremely unnatural and
fine-tuned.  A very reasonable 
stand might be that good taste alone would require us to abandon an
inflationary origin to the perturbations rather than embrace the
conspiratorial models we study here.  The main point of this work is
that modern datasets will leave nothing to good taste.  By the time the
new data is in, even the exotic part of model space we explore here
will be exposed to confrontation with observations.

\section*{Acknowledgements}
We thank Ben Wandelt, Jochen Weller, Lloyd Knox and Hume Feldman for
helpful discussions and Lloyd Knox, Andrew Hamilton and Max Tegmark for 
providing us with data points.  This work was funded in part by PPARC, 
DOE grant DE-FG03-91ER40674 and UC Davis.


\end{document}